\documentclass[12pt,preprint]{aastex}

\usepackage{graphicx}

\begin{document}

\title{OBSERVATIONAL EVIDENCE OF ACCRETION DISK-CAUSED JET PRECESSION IN GALACTIC NUCLEI}

\author{Ju-Fu Lu \textrm{\tiny{AND}} Bo-Yan Zhou}
\affil{Department of Physics, Xiamen University, Xiamen, Fujian 361005,China}
\email{lujf@xmu.edu.cn}

\begin{abstract}
We show that the observational data of extragalactic radio sources tend to
support the theoretical relationship between the jet precession period and the
optical luminosity of the sources, as predicted by the model in which an
accretion disk causes the central black hole to precess.
\end{abstract}


\setlength{\parindent}{.25in}

\section{INTRODUCTION}
An S-(or Z-) shaped morphological symmetry has been observed for many
extragalactic radio sources, for example, for 39 of 365 extended radio sources
collected by Florido, Battaner, \& Sanchez-Saavedra (1990), for more than 30\%
of quasars with redshifts less than 1 and with a two-sided radio structure
(Hutchings, Price, \& Gower 1988). This phenomenon has generally been attributed
to precession in the orientation of the radio jet. With precessing jet models,
it is indeed possible to fit apparently more complex image maps of many extended
radio sources (e.g. Gower et al. 1982). Jet precession was also invoked to
account for other phenomena, such as the bending and misalignment of parsec
scale jets in compact radio sources (e.g. Linfield 1981; Appl, Sol, \& Vicente
1996), and the variability of either continuum or line emissions observed in
active galactic nuclei (e.g. Rieger 2004; Caproni, Mosquera Cuesta, \& Abraham
2004; but see \S 4).

Presumably, the jet material is ejected along the axis of a spinning black hole
at the center of a galactic nucleus. There have been basically two models on
what causes the central black hole to precess. Begelman, Blandford, \& Rees
(1980) first proposed that the precession of the black hole could be due to the
existence of another hole in the same nucleus. Later, Lu (1990) suggested that a
surrounding, titled accretion disk could also do the job, on the basis of the
model of Sarazin, Begelman, \& Hatchett (1980) for the galactic object SS433.
Both the models predicted that there should be a tendency for the brighter
sources to present the shorter precession periods, but the predicted precession
period -- luminosity relations in the two models were quantitatively different
(Roos 1988; Lu 1990). While the idea of binary black holes is indeed interesting
and has been applied in recent years to a number of extragalactic sources with
precessing jets (e.g. Lobanov \& Roland 2005 and references therein), the
scenario of a single black hole with an accretion disk surrounded seems to be
simpler and more commonly applicable to active nuclei. In fact, the disk-caused
jet precession model has received meaningful observational supports since it
arose. For instance, Veilleux, Tully, \& Bland-Hawthorn (1993) reported their
detailed observational results of the Seyfert galaxy NGC 3516, and this source
follows the theoretical precession period -- luminosity relation of Lu (1990);
Peck \& Taylor (2001) revealed that the compact source 1946+708 has radio jets
with a tilted, circumnuclear disk, and stated that such an observed morphology
is consistent with the scenario of Lu (1990).

Most previous studies of jet precession in galactic nuclei were for individual
sources. In this Letter we try to do the job in a more general sense. We first
describe the model of jet precession caused by an accretion disk, which is an
improvement of that proposed by Lu (1990); and then collect all extragalactic
sources to date with needed observational data available to test our model.

\section{THE MODEL}

A spinning (Kerr) black hole has angular momentum $J_* = aGM^2 / c$, where $M$
is the black hole mass, and $a$ is the dimensionless specific angular momentum,
$0 < a \leqslant 1$. Consider an accretion disk surrounding the hole. The disk
is tilted in the sense that its rotation axis is misaligned with that of the
hole. A ring with width $dr$ in the disk has angular momentum $ dJ = 2 \pi r^2
\Sigma v_\phi dr$, where $\Sigma$ is the surface density of the disk, and
$v_\phi$ is the rotational velocity of the disk material, so the angular
momentum per logarithmic interval of radius is $J(r) = dJ / d (\ln r) = 2\pi r^3
\Sigma v_\phi$. There exists a critical radius $r_p$ in the disk defined by $J
(r_p) = J_*$. Due to the Lense-Thirring effect, disk material interior to $r_p$
will be aligned with the equatorial plane of the black hole, while the outer
portion of the disk ($r \gtrsim r_p$) with sufficiently large angular momentum
will maintain its orientation, and will cause the central hole along with the
inner disk to precess (Bardeen \& Petterson 1975). Some material of the inner
disk is assumed to be ejected out along the hole's spin axis, forming a
precessing jet.

The precession rate of the central black hole and the inner disk is equal to
$2GJ(r) /c^2 r^3$ (Sarazin et al. 1980). Since only regions $r \gtrsim r_p$ in
the disk contribute to this precession and $J(r)/r^3$ decreases as $r$
increases, the fastest precession rate $\Omega$ is produced by the ring at $r =
r_p$ with $J(r) = J_*$ , i.e. $\Omega = 2GJ_* /c^2r_p^3$. Introducing the mass
accretion rate $\dot{M}= 2\pi r\Sigma v_r$, where $v_r$ is the radial inflow
velocity of the disk material, $\Omega$ can be expressed as

\begin{equation}
\Omega = (2/M)(G/a)^{1/2}(\dot{M}v_\phi / cv_r)^{3/2}.
\end{equation}

The best known model of accretion disks is that of Shakura \& Sunyaev (1973), in
which $v_\phi$ is Keplerian, $v_\phi = (GM / r)^{1/2}$, and for the outer region
of the disk \\
$v_r = 10^{5.73} \alpha^{ 4/5} m^{-1/5} \dot{m}^{3/10} r_*^{-1/4}
f^{-7/10} \mathrm{cm\ s^{-1}}$, where $\alpha$ is the dimensionless viscosity
parameter, $m = M / M_\sun$, $\dot{m} = \dot{M}/\dot{M}_\mathrm{crit}$ , with
$\dot{M}_\mathrm{crit}$ being the critical accretion rate corresponding to the
Eddington luminosity, $r_* = r / r_g$ , with $r_g = 2GM / c^2$ being the
gravitational radius, and $f = 1 - (3 / r_*)^{1/2} \thickapprox 1$ (Kato, Fukue,
\& Mineshige 1998, p. 88). Substituting the expressions of $v_\phi$ and $v_r$
into equation (1), the precession period $P(\equiv 2\pi/ \Omega)$ can be
calculated numerically,

\begin{equation}
P = 10^{9.25} \alpha^{48/35} a^{5/7} (M/10^8 M_\sun)^{1/7} (\dot{M}/10^{-2}
M_\sun \ \mathrm{yr^{- 1}})^{-6/5} \mathrm{yr}.
\end{equation}

The mass accretion rate $\dot{M}$ is related to the luminosity of the source $L$
as $L = \eta \dot{M} c^2$, with $\eta$ being the efficiency of energy
conversion. Note that here $L$ should be the optical luminosity of the galactic
nucleus, rather than the radio luminosity of the jet, because the jet material
is likely to derive its energy from the spinning black hole or from the disk via
electro-magnetic processes, the jet luminosity does not reflect directly the
accretion rate. By using the definition $\log (L / L_\sun) = 0.4 (5.71 -
M_\mathrm{abs})$, where $M_\mathrm{abs}$ is the absolute magnitude of the
nucleus in the $B$ band, and $5.71$ is $M_B$ of the sun, a relationship between
$P$ and $M_\mathrm{abs}$ is obtained finally,

\begin{equation}
\log P \mathrm{(yr)} = 0.48M_\mathrm{abs} + 20.06 + \frac{48}{35} \log \alpha +
\frac{5}{7} \log a + \frac{1}{7} \log (\frac{M}{10^{8}M_\sun}) + \frac{6}{5}
\log \eta.
\end{equation}
This equation improves significantly equation (5) of Lu (1990), which is $\log P
\mathrm{(yr)} = 0.6 M_\mathrm{abs} + \mathrm{const}.$, as not only the slope
$0.6$ has been recalculated to be $0.48$, but also the previously unclear
constant in the formula has been explicitly expressed.

The critical radius $r_p$ in the disk can be calculated as

\begin{equation}
\begin{array}{ccl}
r_p & = & (J_* v_r/\dot{M} v_\phi)^{1/2}\\
  & = & 10^{17.22} (\alpha/0.1)^{16/35} a^{4/7}(M/10^8 M_\sun)^{5/7} ( \dot{M}/10^{-2} M_\sun \ yr^{- 1})^{-2/5}
  \mathrm{cm},
\end{array}
\end{equation}
which is about several thousand times of the black hole's gravitational radius
$r_g$ . A similar result for $r_p$ has been obtained in the model for accretion
disks in active galactic nuclei given by Collin-Souffrin \& Dumont (1990).

\section{OBSERVATIONAL TESTS}

The theoretical $P-M_\mathrm{abs}$ relationship of equation (3) is practically
testable, as $P$ and $M_\mathrm{abs}$ are observational quantities, and the
ranges of values of the four parameters are all known, namely $ \alpha \thicksim
0.001 - 1$, $0 < a \leqslant 1$, $M \thicksim 10^6 - 10^{10} M_\sun$, and $\eta
\thicksim 0.1$. As mentioned in \S1, many extragalactic radio sources show
phenomena which are suggestive of jet precession. Unfortunately, for most of
those sources no precession periods have been measured, the number of sources
with both the jet precession period and the nucleus magnitude data available is
still rather small. We collect from the literature 41 such sources to our
knowledge, and list them in Table 1. The following comments are in order.
Concerning the precession period, it is usually too long to be observed
directly. The best way known to evaluate this period for a source is by fitting
its (S-shaped or apparently more complex) radio map with a precessing jet model,
like what was done by Gower et al. (1982). This is indeed the case for most
sources listed in Table 1. Not surprisingly, the precession period obtained this
way is by no means definite and accurate, it should be regarded as estimation of
the order of magnitude only. For a few cases (3C 196, 3C 305, M 84, and OJ 287)
the method of map fitting could not be (or has not yet been) used, the
precession period was estimated by some other arguments. As to the sources'
optical absolute magnitudes, they are in general quite inaccurate and
inhomogeneous in the literature. For most of our collected sources we have made
use of the LEDA Database and the SIMBAD Database, where the $M_B$ values can
either be found directly or be calculated from given visual magnitudes $m_B$ and
redshifts $z$. We have made corrections so that all magnitudes in Table 1 are in
the $B$ band, and with $H_0 = 50\ \mathrm{km\ s^{-1} Mpc^{-1}}, q_0 = 0$.

With all these uncertainties in mind, it seems that the data in Table 1 tend to
support an inverse correlation between the jet precession period and the optical
luminosity of the galactic nucleus as equation (3) predicted, i.e. the brighter
the nucleus is, the faster the jet precesses, as shown more clearly in Figure 1.
For a quantitative comparison, we draw in the figure the theoretical $P -
M_\mathrm{abs}$ correlation of equation (3) by three straight lines. The middle
solid line corresponds to the most typical values of the four constant
parameters, i.e. $ \alpha = 0.1$, $a = 0.8$,  $M = 10^8M_\sun$, and $\eta =
0.1$. This line fits the observational data quite well. The upper and the lower
dashed lines are also for $\eta = 0.1$, but for larger values $\alpha = 1$, $a =
1$, and $M = 10^{10} M_\sun$; and smaller values $\alpha = 0.003, a = 0.5$, and
$M = 10^6 M_\sun$, respectively. It is seen that all the 41 sources are located
within the region bounded by these two reasonable limit lines. Thus at this
point the observational data are in good agreement with the theoretical
prediction of the model.

\section{DISCUSSION}

\emph{Precession cone opening angle.} Another physical component of the jet
precession model is the half-opening angle $\psi$ of the precession cone along
which the black hole's spin axis rotates. Unlike the precession period $P$,
$\psi$ has not been calculated and predicted theoretically in our model, nor in
the binary black hole model. The only system in which $\psi$ is known ($\sim
20\degr\ $ ) by observations is the galactic object SS433. For extragalactic
sources, the only way known to evaluate $\psi$ is by fitting a precessing jet
model to the observed morphology and kinematics of the jet. In this fitting,
$\psi$ is deduced as a free parameter along with others such as the precession
period, viewing angle, and speed of the jet (e.g. the early work Gower et al.
1982 that ignored the physical cause of jet precession, and the most recent one
Lobanov \& Roland 2005 that adopted the binary black hole model). Table 1 lists
values of $\psi$ available in the literature, all obtained this way. For the
inner disk with angular momentum much smaller than that of the black hole, one
can expect that the hole's spin axis almost coincides with the total angular
momentum vector along which both the hole and the disk precess, i.e. the hole's
precession cone opening angle is very small (or the hole even does not precess
at all). But for the case we consider here, the outer disk has angular momentum
comparable to or larger than that of the hole, and gas accretion is likely to
occur at random angles, thus initially the two angular momentum vectors can make
any angle, i.e. $\psi$ can be of any value. In a very recent paper, King et al.
(2005; most clearly Figure 1 there) showed that, whatever the initial angle
between the two angular momentum vectors (the hole's and the disk's) is, the
hole suffers a torque which always acts to align the hole's spin with the total
angular momentum, i.e. $\psi$ is always made to decrease with time. It is seen
from Table 1 that the values of $\psi$ are indeed random, and 20 of the total 33
values there are relatively small (1$ \degr$ -- 16$ \degr$). A small $\psi$
could be a feature of the initial configuration of a source, or it could be a
result of evolution from a larger initial $\psi$ due to the alignment torque.

Some insight about $\psi$ may be gained from the study of a test particle
orbiting a black hole. The off-equatorial plane motion of a test particle is
characterized by the Carter constant $Q$. To very high accuracy, $Q$ can be
treated as the square of the particle's specific (per unit mass) angular
momentum projected into the equatorial plane (perpendicular to the hole's spin).
The inclination angle $i$ of the particle's circular orbit is defined as $\cos i
= L_z / (L_z^2 + Q)^{1/2}$, where $L_z$ is the particle's specific angular
momentum parallel to the hole's spin (Hughes \& Blandford 2003). For large $r$,
Hughes (2001) gave a formula (in $G = c = 1$ units): $
(L_z^2+Q)^{1/2}=(rM)^{1/2}[1-3a(M/r)^{3/2} \cos i]$ . It is seen that depending
on the values of two constants $L_z$ and $Q$, $i$ can be of any value (note that
$i$ corresponds to $2\psi$ in our model since we consider a ring with angular
momentum $J$ equal to that of the hole $J_*$), and that whatever the value of
$i$ is, the particle's whole specific angular momentum $(L_z^2 + Q)^{1/2}$
approaches the Keplerian value $(rM)^{1/2}$ as $r$ increases.

\emph{Short term precession?} As seen from equation (2), the jet precession
period is generally very long in our disk-caused precession model. Appl et al.
(1996) have also shown that the typical precession period of a supermassive
black hole caused by a tilted massive torus is of the order of $10^6$ years.
Therefore we include in Table 1 only those sources with precession periods
ranging from $\thicksim 10^3$ to $\thicksim 10^8$ years. We notice that for a
number of extragalactic sources the periodicity of about 10 years in the light
curves has been observed, and this phenomenon has also been attributed by some
authors to precession of jets and/or accretion disks (e.g. Caproni et al. 2004).
In our opinion, such a short periodicity is most likely not related to the
precession. It is due to either flaring activity in the disk, or geometry and
non-linear motion of the jet. At least in some cases, setting such a short
precession period has led to unrealistic results in the model fitting. For
example, Caproni \& Abraham (2004) assumed the period of 10.1 years observed in
the $B$-band light curve of 3C 345 to be the jet precession period, and claimed
that 3C 345 has two black holes with masses of $\thicksim 4\times 10^9 M_\sun$
and an orbital separation $\leq 7.3 \times 10^{16}$ cm. Such a separation is
smaller than $60 r_g$, then there would be no room for a stable accretion disk
to exist at all! This is of course not to argue against the binary black hole
model itself. Also applying the binary black hole model to 3C 345, Lobanov \&
Roland (2005) obtained a precession period of 2570 years, black hole masses of
$7.1\times 10^8 M_\sun$, and an orbital separation $\thicksim
10^{18}\mathrm{cm}$ in their model fitting, and these parameters seem to be more
reasonable. Similar problems may exist for objects 3C 120, 3C 273, and OJ 287,
for which precession with periods $\thicksim$ 10 years was also assumed (see
Table 1 of Caproni et al. 2004), while more plausible precession periods have
been deduced by other authors (Table 1 here).

\emph{Summary}. The observational evidence is likely to be in favor of the
disk-caused jet precession model according to the data in Table 1, at least at a
statistical level, but it is still far from reaching a definite conclusion in
the following senses. First, except that it does not require binary supermassive
black holes to present in a large number of active nuclei, our model does not
offer an explanation of the observational data which is conceptually different
from that offered by the binary black hole model. Some sources, namely 1928+738,
OJ 287, Mrk 501, PKS 0420-014, and 3C 345, though being included in Table 1 to
support our model, were studied by corresponding authors in the framework of
binary black hole model. Second, that the black hole's precession cone angle can
be of any value should be regarded as a hypothesis. In particular, it is not
clear whether this angle as large as several tens degrees is physically
possible. Further investigations addressing this specific question on a
fundamental basis are required. Third, reasons for a large scale misalignment
between the accretion disk and the black hole are also unclear. The misalignment
may be created through gas accretion which is driven for example by minor
mergers of galaxies with satellite galaxies and is likely to occur at random
angles, then it is still a problem whether and how this misalignment can be
preserved over timescales long enough to be reconciled with the extent of
kpc-scale jets showing clear signs of precession.

We thank the referee for helpful comments. This work was supported by the
National Science Foundation of China under grant 10233030 and 10373011.


\begin{deluxetable}{lrrccr}
\tablecaption{ Extragalactic sources with jet precession period $P$,
half-opening angle of precession cone $\psi$, and optical magnitude
$M_\mathrm{abs}$ data available } \tablehead{ \colhead{Object} & \colhead{
$P$(yr) } & \colhead{\hspace{0.2cm} $\psi$ (degrees) } & \colhead{Ref.$^a$ } &
\colhead{ $- M_\mathrm{abs}$ } & \colhead{\hspace{0.5cm} Ref.$^b$}} \startdata
3C 196 & $10^6$ & --- & (1) & 26.4 & (34,35)\\
3C 305 & $5\ 10^6$ & --- & (1) & 22.8 & (34)\\
3C 273 & $ \thicksim 10^3$ & 1 $\thicksim$ 2 & (2) & 27 & (34)\\
3C 129 & $9.2\ 10^6$ & 12  & (3) & 23.3 & (34)\\
3C 315 & $ \thicksim 10^7$ & $<$17 $\pm$ 15  & (4) & 22.6 & (34)\\
3C 388 & $5\ 10^5$ & 6 (to 12 ) & (4) & 23.2 & (34)\\
NGC 326 & $10^6$ & $<$ 23 $\pm$ 3 & (4) & 23.3 & (34)\\
4C 18.68(2305+18) & $4\ 10^4$ & $<$ 80 $\pm$ 6  & (4) & 24.6 & (34,35)\\
1315+347 & $9\ 10^3$ & $<$ 25 $\pm$ 4  & (4) & 25.4 & (35)\\
1730-130 & $5\ 10^4$ & $(<)$ 35  & (4) & 27 & (34,35)\\
0945+408 & $1.6\ 10^4$ & 23 $\pm$ 11  & (4) & 26.9 & (34,36)\\
0224+671 & $3\ 10^4$ & 14 (to 34 ) & (4) & 23.1 & (35)\\
0716+714 & $\thicksim 10^5$ & 35 $\pm$ 8 & (4) & 25.9 & (35)\\
0707+476 & $\thicksim 10^5$ & $(<)$ 4  & (4) & 26.6 & (37)\\
3C 449 & $1.1\ 10^5$ &  $\leqslant$  16  & (5) & 21.3 & (34)\\
1857+566 & $10^4$ & 5  & (6) & 27.3 & (34,35)\\
4C 29.47 & $6\ 10^6$ & 60 $\pm$ 10  & (7) & 21.9 & (34)\\
2300-189 & $3.4\ 10^6$ & 69 $\pm$ 5  & (8) & 21.8 & (34)\\
3C 418 & $\thicksim 5\ 10^4$ & $\thicksim$70  & (9) & 29.1 & (34,35)\\
3C 120 & $5\ 10^4$ & 9  & (10) & 22.9 & (34)\\
NGC 6251 & $1.8\ 10^6$ & 8  & (11) & 22.5 & (34)\\
4CT74.17.1 & $9\ 10^5$ & 10  & (12) & 24.4 & (38)\\
3C 138 & $8200$ & 1  & (13) & 24.7 & (34,36)\\
Hydra A (3C 218) & $3\ 10^5$ & --- & (14) & 23.8 & (34)\\
4C 41.17 & $3\ 10^6$ & --- & (15) & 25.7 & (39)\\
3C 119 & $2\ 10^3$ or $6\ 10^4$ & 3 or 17.5 & (16) & 26.4 & (34,35)\\
1928+738 & $4\ 10^5$ & 8 $\thicksim$ 10  & (17) & 26.3 & (34,36)\\
NGC 3516 & $\thicksim 10^7$ & 60  & (18) & 21.6 & (34)\\
B2 0828+32 & $2\ 10^8$ & 25  & (19,20) & 20.7 & (34)\\
NGC 3079 & $1.2\ 10^6$ & --- & (21) & 22.1 & (34)\\
3C 390.3 & $4\ 10^5$ & --- & (22) & 21.2 & (34)\\
1243+036 & $3.6\ 10^5$ & --- & (23) & 25.4 & (39)\\
OJ 287 & $7\ 10^4$ & 3.4  & (24) & 25.5 & (37)\\
0153+744 & $880$ & 10  & (25) & 29.4 & (37)\\
NGC 5128(Cen A) & $\thicksim10^7$ & --- & (26) & 23 & (34,36)\\
M 84(NGC 4374) & $ \thicksim 10^7$ & 10--15  & (27) & 21.6 & (34)\\
Mrk 501(1652+398) & $ \thicksim 10^4$ & $<$ 12  & (28,29) & 23.5 & (34)\\
NGC 4258 & $2.5\ 10^5$ & several tens  & (30) & 21.8 & (34)\\
PKS 0420-014 & $\thicksim 10^4$ & $\sim$ 12  & (31) & 27.3 & (34,35)\\
3C 345 & $2570$ & 1.45  & (32) & 27.1 & (34,35)\\
PKS 2153-69 & $1.8\ 10^6$ & 11.6  & (33) & 21.9 & (34)\\
\enddata
\begin{flushleft}\hspace{0.5cm} $^a$ References for $P$ and $\psi$ ;  $^b$ References for $M_\mathrm{abs}$. \end{flushleft}
\tablerefs{(1) Lonsdale \& Morison 1980; (2) Linfield 1981; (3) Icke 1981;
 (4) Gower et al. 1982; (5) Gower \& Hutchings 1982; (6) Saikia et al. 1983; (7) Condon \& Mitchell
 1984;
(8) Hunstead et al. 1984; (9) Muxlow et al. 1984; (10) Benson et al. 1984; (11)
Jones et al. 1986; (12) Roos \& Meurs 1987; (13) Fanti et al. 1989; (14) Taylor
et al. 1990; (15) Chambers et al. 1990; (16) Nan et al. 1991; (17) Hummel et al.
1992; (18) Veilleux et al. 1993; (19) Klein et al. 1995; (20) Parma et al.1985;
(21) Baan \& Irwin 1995; (22) Gaskell 1996; (23) van Ojik et al. 1996; (24)
Vicente et al. 1996; (25) Hummel et al. 1997; (26) Israel 1998; (27) Quillen \&
Bower 1999; (28) Rieger \& Mannheim 2000; (29) Conway \& Wrobel 1995; (30) Cecil
et al. 2000; (31) Britzen et al. 2001; (32) Lobanov \& Roland 2005; (33) Young
et al. 2005; (34) LEDA Database; (35) SIMBAD Database; (36) NASA/IPAC
Extragalactic Database; (37) V\'{e}ron-Cetty \& V\'{e}ron 2003; (38) Wirth et
al. 1982; (39) van Breugel et al. 1998.}
\end{deluxetable}

\clearpage

\begin{figure}
\plotone{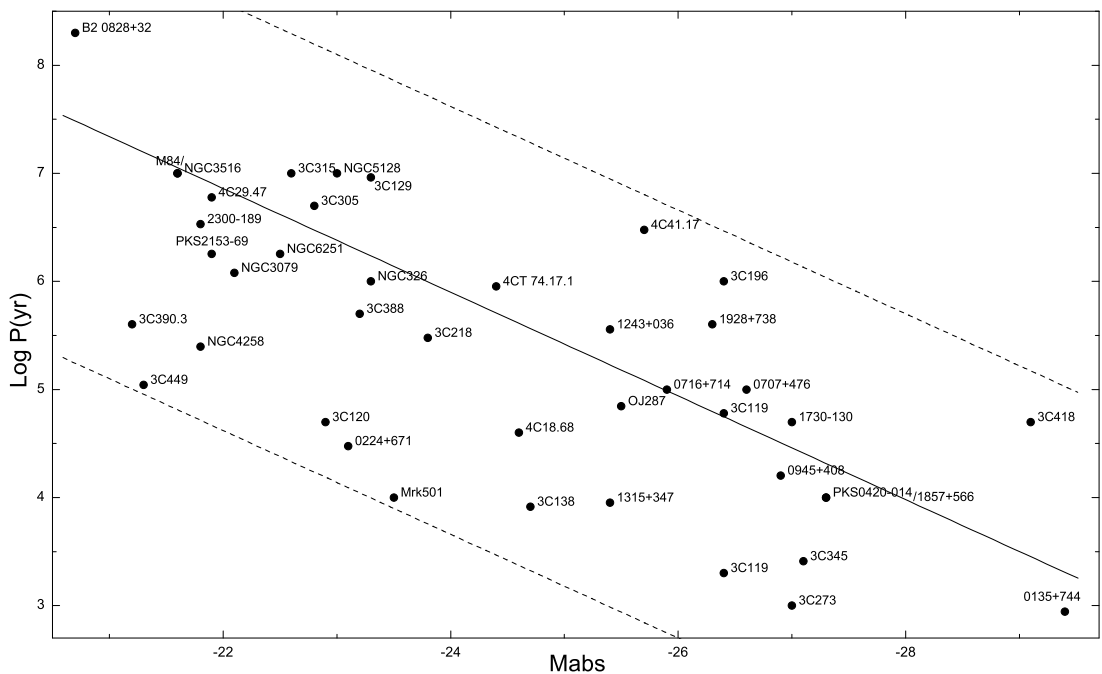} \caption{Precession periods $P$ vs. optical absolute magnitudes
$M_\mathrm{abs}$ for 41 extragalactic sources listed in Table 1. Three straight
lines draw the theoretical prediction equation (3), the upper, middle, and lower
ones are for parameters ($ \alpha , a, M/M_\sun, \eta $) equal to ($1, 1,
10^{10}, 0.1), (0.1, 0.8, 10^8, 0.1)$, and ($0.003, 0.5, 10^6, 0.1$),
respectively.}
\end{figure}

\end{document}